\documentclass[11pt]{article}
\usepackage[utf8]{inputenc}
\usepackage[english]{babel}
\usepackage[active]{srcltx}
\usepackage{hyperref}
\usepackage[all]{xy}
\usepackage{amsfonts}
\usepackage{amsthm}
\usepackage{enumerate}
\usepackage{mathrsfs}
\usepackage{multicol}
\usepackage[all]{xy}
\usepackage{amsfonts}
\usepackage{latexsym,amsmath,amssymb}
\usepackage[usenames,dvipsnames]{color}
\usepackage{xfrac}
\usepackage{pstricks}
\usepackage[absolute,overlay]{textpos}
\usepackage{graphics,wrapfig,times}

\usepackage{color}
\definecolor{DarkBlue}{rgb}{0.1,0.1,0.5}
\definecolor{Red}{rgb}{0.9,0.1,0.1}
\definecolor{Green}{rgb}{0.3,0.7,0.0}
\definecolor{green2}{rgb}{0.1,0.7,0.2}
\definecolor{blue2}{rgb}{0.0,0.6,0.7}
\definecolor{pink}{rgb}{1,0.0,1}
\definecolor{orange}{rgb}{0.9,0.0,0.1}

\newtheorem{theo}{Theorem}

\newtheorem{prop}{Proposition}

\newtheorem{definition}{Definition}
\newtheorem{remark}{Remark}

\renewcommand{\d}{\mathrm{d}}
\newcommand{\derpar}[2]{\displaystyle\frac{\partial{#1}}{\partial{#2}}}

\newcommand{\vf}{\mathfrak{X}}
\newcommand{\df}{\Omega}

\newcommand{\Tan}{\mathrm{T}}
\newcommand{\inn}{{\mathop{i}\nolimits}}
\newcommand{\Lie}{\mathop{\mathrm{L}}\nolimits}

\newcommand{\bal}{\begin{align*}}
\newcommand{\eal}{\end{align*}}
\def\beq{\begin{equation}}
\def\eeq{\end{equation}}
\def\bea{\begin{eqnarray}}
\def\eea{\end{eqnarray}}
\def\beann{\begin{eqnarray*}}
\def\eeann{\end{eqnarray*}}
\def\ben{\begin{enumerate}}
\def\een{\end{enumerate}}
\def\bit{\begin{itemize}}
\def\eit{\end{itemize}}

\def\vf{\mathfrak X}
\def\df{{\mit\Omega}}

\def\d{{\rm d}}

\def\Real{\mathbb{R}}

\def\Tan{{\rm T}}
\def\Lie{\mathop{\rm L}\nolimits}
\def\inn{\mathop{i}\nolimits}
\def\Cinfty{{\rm C}^\infty}

\def\tabaddress#1{{\small\it\begin{tabular}[t]{c}#1
\\[1.2ex]\end{tabular}}}

\title{\sc Some properties of Multisymplectic Manifolds}
\author{
   {\sc Narciso Rom\'an-Roy\thanks{{\bf e}-{\it mail}:
   narciso.roman@upc.edu  / ORCID: 0000-0003-3663-9861.}}  \\
   \tabaddress{Department of Mathematics.
   Ed. C-3, Campus Norte UPC\\
   C/ Jordi Girona 1. 08034 Barcelona. Spain.}}
   \date{\today \\
   }

\begin{document}

\maketitle

\pagestyle{myheadings}
\markright{\rm N. Rom\'an-Roy,
   \sl Some properties of multisymplectic manifolds}
\maketitle
\thispagestyle{empty}

\begin{abstract}
This lecture is devoted to review some of the main properties of multisymplectic geometry.
In particular, after reminding the standard definition of multisymplectic manifold,
we introduce its characteristic submanifolds, the canonical models,
and other relevant kinds of multisymplectic manifolds, such as those where
the existence of Darboux-type coordinates is assured.
The Hamiltonian structures that can be defined in these manifolds are also studied, as well as other important properties, such as
their invariant forms and the characterization by automorphisms.
\end{abstract}


\bigskip
\vbox{\raggedleft AMS s.\,c.\,(2010): \null 
{\it Primary}: 53D42, 55R10, 70S05, 83C05; {\it Secondary}: 49S05, 53C15, 53C80, 53Z05.}\null


\section{Introduction}
\label{intro}

Although there are several geometrical models for describing classical field theories, namely,
{\sl polysymplectic, $k$-symplectic} and  {\sl $k$-cosymplectic manifolds} \cite{LSV-2016,GMS-97,Ka-98,RRS-2005,RRSV-2011};
{\sl multisymplectic manifolds}
are the most general and complete tool for describing geometrically
(covariant) first and higher-order field theories
(see, for instance,\cite{AA-80,CCI-91,LMM-96,EMR-96,Gc-73,GS-73,GIM1,KT-79,pere,art:Roman09,book:Saunders89} and the references quoted on them).
All of these kinds of manifolds are generalizations of the concept of
{\sl symplectic manifold},
which are used to describe geometrically mechanical (autonomous) systems.

This talk is devoted to review some of the main properties of multisymplectic geometry
and is mainly based on the results presented in
\cite{Ca96a,Ca96b,LMS-2003,EIMR-2012,Ib-2000}.
In particular we discuss the following topics:
the basic definition of {\sl multisymplectic manifold}
(in Section 2) and the {\sl Hamiltonian structures}
associated to a multisymplectic form (Section 3),
the characteristic submanifolds of multisymplectic manifolds (Section 4),
the canonical models and the existence of Darboux-type coordinates
(Section 5),
other kinds of relevant multisymplectic manifolds (Section 6) and,
finally,
some interesting theorems of invariance and characterization
by automorphisms (Section 7).

All the manifolds are real, second countable and $\Cinfty$. The maps and the structures are $\Cinfty$.  Sum over repeated indices is understood.


  \section{Multisymplectic manifolds}

(See \cite{Ca96a,Ca96b,EIMR-2012} for more details).

\begin{definition}
Let $M$ be a differentiable manifold, 
with $\dim M=n$,
and $\Omega\in\df^k(M)$ ($\df^k(M)$ denotes the set of differentiable $k$-forms in $M$), with $k\leq n$.
\begin{itemize}
\item
The form $\Omega$ is {\rm 1-nondegenerate} if, for every $p \in M$
and $X_p\in\Tan_pM$,
$$
\inn(X_p) \Omega_p = 0
\ \Longleftrightarrow \ X_p = 0 \ .
$$
\item
The form $\Omega$ is a {\rm multisymplectic form}
 if it is closed and $1$-nondegenerate.
\item
A {\rm multisymplectic manifold} of  {\rm degree $k$}
is a couple $(M,\Omega)$, where
$\Omega\in\df^k(M)$ is a multisymplectic form.
\end{itemize}

If $\Omega$ is only closed then it is called a
{\rm pre-multisymplectic form}.

If $\Omega$ is only $1$-nondegenerate then it is an
{\rm almost-multisymplectic form}.
\end{definition}

If $\dim M \geq 2$, then a multisymplectic $k$-form must have degree $k \geq 2$.

The property of $1$-nondegeneracy can be characterized equivalently as follows:
a differentiable $k$-form $\Omega$ is 1-nondegenerate if, and only if, 
the vector bundle morphism
\[
\begin{array}{rrcl}
\Omega^\flat \colon & \Tan M & \to & \mathsf{\Lambda}^{k-1} \Tan^*M
\\
& X_p & \mapsto & \inn(X_p)\Omega_p
\end{array}
\]
and thus the corresponding morphism of $\mathcal{C}^\infty(M)$-modules
\[
\begin{array}{rrcl}
\Omega^\flat \colon & \mathfrak{X}(N)& \to & \Omega^{k-1}(N)
\\
& X & \mapsto & \inn(X)\Omega
\end{array}
\]
are injective.

Some examples of multisymplectic manifolds are the following:
Multisymplectic manifolds of degree $2$ are just {\sl symplectic manifolds}.
Multisymplectic manifolds of degree $n$ are {\sl orientable manifolds} 
and the multisymplectic forms are {\sl volume forms}.
{\sl Bundles of $k$-forms ($k$-multicotangent bundles)}
endowed with their {\sl canonical $(k+1)$-forms}
are multisymplectic manifolds of degree $k+1$.
{\sl Jet bundles} (over $m$-dimensional manifolds) 
endowed with the {\sl Poincar\'e-Cartan $(m+1)$-forms}
associated with  (singular){\sl Lagrangian densities}
are (pre)multisymplectic manifolds of degree $m+1$.


  \section{Hamiltonian structures in multisymplectic manifolds}
\label{sec3}

(See \cite{Ca96a,Ca96b,EIMR-2012} for more details).

\begin{definition}
A { \rm $m$-vector field} (or a { \rm multivector field of degree $m$}) 
in a manifold $M$ (with $m\leq n=\dim M$)
is any section of the bundle $\Lambda^m(\Tan M)\to M$
(that is, a contravariant, skewsymmetric
tensor field of degree $m$ in $M$).
The set of $m$-vector fields in $M$ is denoted by $\vf^m (M)$.
\end{definition}

The local description of multivector fields of degree $m$ is the following:
for every $p\in M$,  there are a neighbourhood $U_p\subset M$ and local vector fields
$X_1,\ldots ,X_r\in\vf (U_p)$, with $m\leq r\leq{\rm dim}\, M$, such that
\beq
{\bf X}\vert_{U_p}=\sum_{1\leq i_1<\ldots <i_m\leq r}
f^{i_1\ldots i_m}X_{i_1}\wedge\ldots\wedge X_{i_m}
\ ; \ \mbox{\rm with $f^{i_1\ldots i_m}\in\Cinfty (U_p)$}\ .
\label{1}
\eeq

\begin{definition}
Let ${\bf X}\in\vf^m(M)$ be a multivector field.

\noindent ${\bf X}$ is { \rm homogeneous} (or { \rm  decomposable}) 
if there are $X_1,\ldots ,X_m\in\vf (M)$ such that
${\bf X}=X_1\wedge\ldots\wedge X_m$.

\noindent ${\bf X}$ is { \rm locally homogeneous (decomposable)} if,
for every $p\in M$, there exist $ U_p\subset M$
and $X_1,\ldots ,X_m\in\vf (U_p)$ such that
${\bf X}\vert_{U_p}=X_1\wedge\ldots\wedge X_m$.
\end{definition}

\begin{remark}
{\rm Locally decomposable $m$-multivector fields ${\bf X}\in\vf^m(M)$ 
are locally associated with $m$-dimensional
distributions $D\subset{\rm T}M$.}
\end{remark}

Every multivector field ${\bf X}\in\vf^m(M)$ defines a 
{\sl contraction} with differential forms $\Omega\in\df^k(M)$,
which is the natural contraction between tensor fields.
In particular, if ${\bf X}$ is expressed as in \eqref{1}, we have
\beann
\inn({\bf X})\Omega\vert_{U_p}&=&
\sum_{1\leq i_1<\ldots <i_m\leq r}f^{i_1\ldots i_m}
\inn(X_1\wedge\ldots\wedge X_m)\Omega 
\\ &=&
\sum_{1\leq i_1<\ldots <i_m\leq r}f^{i_1\ldots i_m}
\inn (X_1)\ldots\inn (X_m)\Omega \ .
\eeann
Then, the $k$-form $\Omega$ is said to be {\sl $j$-nondegenerate} 
(for $1\leq j\leq k-1$) if, for every $p\in E$ and ${\bf Y}\in\vf^j(M)$, 
we have that $\inn({\bf Y}_p)\Omega_p =0$ if, and only if, ${\bf Y}_p=0$.

Then, for every  form $\Omega\in\df^k(M)$ ($k\geq m$) we have the morphisms
$$
\begin{array}{cccccc}
\Omega^\flat &\colon&\Lambda^m (\Tan M) &
\longrightarrow & \Lambda^{k-m}(\Tan^* M) &
\\
& & {\bf X}_p & \mapsto & \inn({\bf X}_p)\Omega_p &
\end{array}
\ ; \
\begin{array}{cccccc}
\Omega^\flat &\colon&\vf^m (M) &
\longrightarrow & \df^{k-m}(M) &
\\
& & {\bf X} & \mapsto & \inn({\bf X})\Omega & .
\end{array}
$$

In addition, if ${\bf X}\in\vf^m(M)$, the {\sl Lie derivative} of $\Omega\in\df^k(M)$ is
$$
\Lie({\bf X})\Omega:=[\d , \inn ({\bf X})]\Omega=
\d\inn ({\bf X})\Omega-(-1)^m\inn ({\bf X})\d\Omega \ .
$$

\begin{definition}
Let $(M,\Omega)$ be a multisymplectic manifold of degree $k$.
A diffeomorphism $\varphi\colon M\to M$ is a 
{\rm multisymplectomorphism}
if $\varphi^*\Omega = \Omega$. 
\end{definition}

\begin{definition}
Let $(M,\Omega)$ be a multisymplectic manifold of degree $k$.
\ben
\item
A vector field $X\in\vf(M)$ is a {\rm locally Hamiltonian vector field} if its flow consists of multisymplectic diffeomorphisms. 
It is equivalent to demand that
$\Lie (X) \Omega = 0$, or equivalently,
 $\inn(X) \Omega\in\df^{k-1}(M)$ is a closed form.
\item
A multivector field
${\bf X}\in\vf^m(M)$ ($m<k$) is a {\rm locally Hamiltonian multivector field}
if $\Lie ({\bf X})\Omega=0$ or, what is equivalent,
$\inn ({\bf X})\Omega\in\df^{k-m}(M)$ is a closed form.
Then, for every $p\in M$, exist $U\subset M$ and $\zeta \in\df^{k-m-1}(U)$ such that
$\inn ({\bf X})\Omega = \d\zeta$ (on\ $U$).

In this case $\zeta\in\df^{k-m-1}(U)$ is said to be a {\rm locally Hamiltonian form} for ${\bf X}$.
\item
${\bf X}\in\vf^m(M)$ is a {\rm Hamiltonian multivector field}
if $\inn ({\bf X})\Omega\in\df^{k-m}(M)$ is an exact form; that is,
there exists $\zeta \in\df^{k-m-1}(M)$ such that
\ $\inn ({\bf X})\Omega =\d\zeta$.

In this case $\zeta\in\df^{k-m-1}(M)$ is said to be a {\rm Hamiltonian form} for ${\bf X}$.
\een
\end{definition}


 \section{Characteristic submanifolds of multisymplectic manifolds}

(See \cite{Ca96b,LMS-2003} for more details and proofs).

\begin{definition}
Let $(M,\Omega)$ be a multisymplectic manifold of degree $k$,
and ${\mathcal W}$ a distribution in $M$.
$\forall p\in M$ and $1\leq r\leq k-1$,
the {\rm $r$-orthogonal multisymplectic vector space} at $p$ is
$$
{\mathcal W}_p^{\perp,r}=\{ v\in\Tan_pM\,\vert\,
\inn(v\wedge w_1\wedge\ldots\wedge w_r)\Omega_p=0,\
\forall w_1,\ldots,w_r\in{\mathcal W}_p\} \ ,
$$
the {\rm $r$-orthogonal multisymplectic complement} of ${\cal W}$
is the distribution
$\displaystyle{\mathcal W}^{\perp,r}:=\displaystyle\cup_{p\in M}{\mathcal W}_p^{\perp,r}\ .$
\ben
\item
${\mathcal W}$ is an {\rm $r$-coisotropic distribution} if
\ ${\mathcal W}^{\perp,r}\subset{\mathcal W}$.
\item
${\mathcal W}$ is an {\rm $r$-isotropic distribution} if
\ ${\mathcal W}\subset{\mathcal W}^{\perp,r}$.
\item
${\mathcal W}$ is an {\rm $r$-Lagrangian distribution} if
\ ${\mathcal W}={\mathcal W}^{\perp,r}$.
\item
${\mathcal W}$ is a {\rm multisymplectic distribution} if
\ ${\mathcal W}\cap{\mathcal W}^{\perp,k-1}=\{ 0\}$.
\een
\end{definition}

\begin{remark}
{\rm For every distribution ${\cal W}$, we have that ${\cal W}^{\perp,r}\subset{\cal W}^{\perp,r+1}$.
As a consequence,
every $r$-isotropic distribution is $(r+1)$-isotropic, and 
every $r$-coisotropic distribution is $(r-1)$-coisotropic.}
\end{remark}

As a particular situation, if we have a submanifold $N$ of
multisymplectic manifold $M$, we can take as distribution in $\Tan M$ the tangent bundle $\Tan N$ and this allows us to establish 
a classification of these submanifolds as follows:

\begin{definition}
Let $(M,\Omega)$ be a multisymplectic manifold of degree $k$,
and $N$ a submanifold of $M$.
If \ $0\leq r\leq k-1$, then:
\ben
\item
$N$ is an {\rm $r$-coisotropic submanifold} of $M$ if
\ $\Tan N^{\perp,r}\subset\Tan N$.
\item
$N$ is an {\rm $r$-isotropic submanifold} of $M$ if
\ $\Tan N\subset\Tan N^{\perp,r}$.
\item
$N$ is an {\rm $r$-Lagrangian submanifold} of $M$ if
\ $\Tan N=\Tan N^{\perp,r}$.
\item
$N$ is a {\rm multisymplectic submanifold} of $M$ if
\ $\Tan N\cap\Tan N^{\perp,k-1}=\{ 0\}$.
\een
\end{definition}

And, in particular we have:

\begin{prop}
A submanifold $N$ of $M$ is $r$-Lagrangian if, and only if, it is $r$-isotropic and maximal.
\end{prop}


  \section{Canonical models for multisymplectic manifolds. Darboux-type coordinates}

(See \cite{LMS-2003} for more details).

In the same way as the tangent bundle of a manifold is the canonical model for symplectic manifolds,
the canonical models of multisymplectic manifolds are the {\sl bundles of forms}.
These canonical models are constructed as follows:
\bit
\item
If $Q$ is a manifold,
the bundle $\rho\colon\Lambda^{k}(\Tan^*Q)\to Q$
is the {\sl  bundle of $k$-forms} in $Q$.

The { \sl tautological form} (or {\sl canonical form\/})
$\Theta_Q\in\df^{k}(\Lambda^{k}(\Tan^*Q))$
is defined as follows:
if $\alpha\in\Lambda^k(\Tan^*Q)$,
 and $V_1,\ldots,V_k\in\Tan_\alpha(\Lambda^k(\Tan^*Q))$, then
 $$
\Theta_{Q_\alpha}(V_1,\ldots ,V_k)=\inn(\rho_*V_k\wedge\ldots\wedge\rho_* V_1)\alpha \ .
 $$
We have that, $\Omega_Q=\d\Theta_Q\in\df^{k+1}(\Lambda^k(\Tan^*Q))$
is a $1$-nondegenerate form and then
$(\Lambda^k(\Tan^*Q),\Omega_Q)$ is a multisymplectic manifold
of degree $k+1$.

If $(x^i,p_{i_1\ldots i_k})$ is a system of { \sl natural coordinates}
in $U\subset\Lambda^k(\Tan^*Q)$, then the local expressions of these 
canonical forms are
$$
\Theta_Q\mid_U=p_{i_1\ldots i_k}\d x^{i_1}\wedge\ldots\wedge\d x^{i_k}
\quad , \quad
\Omega_Q\mid_U=\d p_{i_1\ldots i_{k}}\wedge\d x^{i_1}\wedge\ldots\wedge\d x^{i_{k}} \ .
$$
These are called { \sl Darboux coordinates} in $\Lambda^k(\Tan^*Q)$.
\item
If $\pi\colon Q\to E$ is a fibration,
let $\rho_r\colon\Lambda^k_r(\Tan^*Q)\to Q$ be the subbundle of $\Lambda^k(\Tan^*Q)$
made of the $r$-horizontal $k$-forms in $Q$ with respect to the projection $\pi$
(that is, the $k$-forms in $Q$ vanishing when applied to $r$ $\pi$-vertical vector fields in $Q$).

Let $\Theta^r_Q\in\df^k(\Lambda^k_r(\Tan^*Q))$ be the pull-back of
$\Theta_Q$ to $\Lambda^k_r(\Tan^*Q)$. This is
the tautological $k$-form in $\Lambda^k_r(\Tan^*Q)$, and then,
if we construct
$\Omega^r_Q=\d\Theta^r_Q\in\df^{k+1}(\Lambda^k_r(\Tan^*Q))$,
we have that
$(\Lambda^k_r(\Tan^*Q),\Omega^r_Q)$ is also a multisymplectic manifold of degree $k+1$.

In the same way, there are also charts of Darboux coordinates
in $\Lambda^k_r(\Tan^*Q)$ on which these canonical forms have a local expressions similar to the above ones.
\eit

Nevertheless, unlike symplectic manifolds,
multisymplectic manifolds $(M,\Omega)$ in general are not (locally)
diffeomorphic to their canonical models, and additional properties are needed
in order to have a { Darboux theorem}
which assures the existence of Darboux-type coordinates
\cite{Mar-88}.
In particular:

\begin{definition}
A {\rm special multisymplectic manifold} is a multisymplectic
manifold $(M,\Omega)$ of degree $k$ such that:
\ben
\item
$\Omega=\d\Theta$, for some $\Theta\in\df^{k-1}(M)$.
\item
There is a diffeomorphism $\phi\colon M\to \Lambda^{k-1}(\Tan^*Q)$,
$\dim\, Q=n\geq k-1$,
(or $\phi\colon M\to \Lambda^{k-1}_r(\Tan^*Q)$),
and a fibration $\pi\colon M\to Q$
such that $\rho\circ\phi=\pi$
(resp. $\rho_r\circ\phi=\pi$),
and $\phi^*\Theta_Q=\Theta$ (resp. $\phi^*\Theta_Q^r=\Theta$).

(It is said that $(M,\Omega)$ is {\rm  multisymplectomorphic to a bundle of forms}).
\een
\end{definition}

In order to have multisymplectic manifolds 
which locally behave as the canonical models, 
it is necessary to endow them with additional structures;
in particular,
a 1-isotropic distribution ${\cal W}$ satisfying some dimensionality conditions, 
and  a ``generalized distribution'' $\varepsilon$ defined
on the space of leaves determined by ${\cal W}$. 
In fact, the existence of distributions satisfying certain properties
is a necessary condition in order to establish Darboux-type theorems for
different kinds of geometrical structures (presymplectic, cosymplectic, $k$-(pre)symplectic, and $k$-(pre)cosymplectic) \cite{Awane-1992,book:Crampin_Pirani,LMS-1988,LMORS-1998,GRR-2018}.
Thus:

\begin{definition}
Let $(M,\Omega)$ be a multisymplectic manifold of degree $k$,
and ${\mathcal W}$ a regular $1$-isotropic involutive distribution in $(M,\Omega)$.
\ben
\item
A {\rm multisymplectic manifold of type $(k,0)$} is a triple 
$(M,\Omega,{\mathcal W})$ such that,
for every $p\in M$, we have that:
\ben
\item
${\rm dim}\,{\mathcal W}(p)={\rm dim}\,\Lambda^{k-1}(\Tan_pM/{\mathcal W}(p))^*$.
\item
${\rm dim}\,(\Tan_pM/{\mathcal W}(p))>k-1$.
\een
\item
A {\rm  multisymplectic manifold of type $(k,r)$}
($1\leq r\leq k-1$) is a quadruple
$(M,\Omega,{\mathcal W},{\mathcal E})$,
where ${\mathcal E}$ is a ``generalized distribution'' on $M$
(this means that, for every $p\in M$,
${\mathcal E}(p)$ is a vector subspace of $\Tan_pM/{\mathcal W}(p)$) 
and, denoting by $\pi_p\colon\Tan_pM\to\Tan_pM/{\mathcal W}(p)$
the canonical projection, we have that:
\ben
\item
$\inn(v_1\wedge\ldots\wedge v_r)\Omega_p=0$, for every $v_i\in\Tan_pM$
such that $\pi_p(v_i)\in{\mathcal E}(p)$ ($i=1,\ldots,r$).
\item
${\rm dim}\,{\mathcal W}(p)={\rm dim}\,\Lambda_r^{k-1}(\Tan_pM/{\mathcal W}(p))^*$,
where the horizontal forms are considered with respect to the
subspace ${\mathcal E}(p)$.
\item
${\rm dim}\,(\Tan_pM/{\mathcal W}(p))>k-1$.
\een
\een
\end{definition}

And the fundamental result is the following:

\begin{prop}
Every multisymplectic manifold $(M,\Omega)$ of type $(k,0)$ 
(resp. of type $(k,r)$)
is locally multisymplectomorphic to a bundle of $(k-1)$-forms
$\Lambda^{k-1}(\Tan^*Q)$ (resp. $\Lambda^{k-1}_r(\Tan^*Q)$),
for some manifold $Q$; that is, to a canonical multisymplectic manifold.

Therefore, there is a local chart of {\sl Darboux coordinates}
 around every point $p\in M$.
\end{prop}
\noindent({\sl Proof\/}):
The proof of this Theorem is very long and can be found in \cite{LMS-2003} (where, in particular, the relation with the canonical models is shown).
\qed

Then we define:

\begin{definition}
Multisymplectic manifolds which are locally multisymplectomorphic
to bundles of forms are called 
{\rm  locally special multisymplectic manifolds}.
\end{definition}

Of course, every special multisymplectic manifold is a locally special multisymplectic manifold and hence
has charts of {\sl Darboux coordinates} at every point.

As an interesting example, if $\pi\colon E\to M$ is a fiber bundle 
(where $M$ is an $m$-dimensional oriented manifold), 
$J^1\pi$ is the corresponding
first-order jet bundle, and ${\cal L}$ is a first-order hyperregular 
Lagrangian density, then the Poincar\'e-Cartan form 
$\Omega_{\cal L}\in\df^{m+1}(J^1\pi)$
is a multisymplectic form and $(J^1\pi,\Omega_{\cal L})$ is a
special multisymplectic manifold
\cite{CCI-91,EMR-00,art:Roman09}.


  \section{Other kinds of multisymplectic manifolds}

(See \cite{EIMR-2012} for more details).

It is a well-known property of symplectic manifolds that the set of
local Hamiltonian vector fields span locally the tangent bundle of the manifold and, hence, the action of the group of multisymplectic diffeomorphisms on $M$ is transitive
(in fact, these properties are a consequence of the existence of Darboux coordinates).
Nevertheless,  in general, these properties do not hold for multisymplectic manifolds and so locally Hamiltonian vector fields
in a multisymplectic manifold $(M,\Omega)$
do not span the tangent bundle of this manifold,
and the group of multisymplectic diffeomorphisms does not act transitively on $M$.
In order to achieve this we need to introduce additional conditions.
Hence, we define:

\begin{definition}
 Let $M$ be a differentiable manifold, $p\in M$ and a compact set $K$
 with $p\in\stackrel{\circ}{K}$.
A {\rm local Liouville} or 
{\rm local Euler-like vector field }at $p$ with respect to $K$ 
is a vector field $\Delta^p\in\vf(M)$ such that:
\ben
\item
${\rm supp}\,\Delta^p:=\overline{\{ x\in M\, | \, \Delta^p(x)\not= 0\}} \subset K$,
\item
there exists a diffeomorphism 
$\varphi \colon\overbrace{{\rm supp}\,\Delta^p}^{\circ} \to \Real^n$
such that $\varphi_*\Delta^p = \Delta$,
where $\displaystyle \Delta=x^i\derpar{}{x^i}$ is the standard Liouville or dilation vector field in $\Real^n$.
\een
\end{definition}

\begin{definition}
 A form $\Omega\in\df^k(M)$ is said to be 
{\rm locally homogeneous} at $p\in M$ if,
for every open set $U\subset M$ containing $p$, 
there exists a local Euler-like vector field $\Delta^p$ at $p$
with respect to a compact set $K\subset U$ such that
$$
\Lie({\Delta^p}) \Omega = f \Omega \ ;\quad f\in\Cinfty(U) \ .
$$
$\Omega$ is {\rm locally homogeneous} if it is locally homogeneous for all $p\in M$.

A {\rm locally homogeneous manifold}
is a couple $(M,\Omega)$, where $M$ is a manifold and
$\Omega\in\df^k(M)$ is locally homogeneous.
 \end{definition}

Therefore we have that:

\begin{prop}
\label{propo3}
Let $(M,\Omega)$ be a locally homogeneous multisymplectic manifold.
Then the family of locally Hamiltonian vector
fields span locally the tangent bundle of $M$; that is, $\forall\, p\in M$,
$
\Tan_p M = {\rm span} \{X_p \mid X\in\vf (M)\ ,\ \Lie (X) \Omega = 0\} \ .
$
\end{prop}
\noindent({\sl Outline of the proof\/}):
The proof is very technical (see  \cite{EIMR-2012} for all the details).
First, the existence of local Euler-like vector fields and their properties
allows us to prove a previous result known as the {\sl localization Lemma} 
which states that, if $X$ is a locally Hamiltonian vector field,
and $x_0\in M$, then for each open set $U\ni x_0$,
there exists an open neighbourhood $V$ of $x_0$ such that $V\subset \bar{V} \subset U$,
with $\bar{V}$ compact, and a locally Hamiltonian vector field $X'$
such that $X'$ coincides with $X$ in $V$ and vanishes identically outside of $U$.
Then, the proof of this Proposition follows from the aplication of this Lemma and using again Euler-like vector fields.
\qed

\begin{theo}
The group of multisymplectic diffeomorphisms 
$G(M,\Omega)$ 
of a locally homogeneous  multisymplectic
manifold $(M,\Omega)$ acts transitively on $M$.
\label{theor1}
\end{theo}
\noindent({\sl Outline of the proof\/} \cite{EIMR-2012}):
The proof is based on the application of Proposition \ref{propo3}
and the above mentioned localization Lemma.
\qed

\begin{remark}
{\rm
Locally special multisymplectic manifolds have local
Euler-like vector fields; in particular, the local vector fields
$\displaystyle \left\{ x^i\derpar{}{x^i}+p_{i_1\ldots i_k}\derpar{}{p_{i_1\ldots i_k}}\right\}$.
Then, the corresponding multisymplectic forms are locally
homogeneous.

As a consequence,
if $(M,\Omega)$ is a locally special multisymplectic manifold,
then the family of locally Hamiltonian vector
fields span locally the tangent bundle of $M$ and
the group of multisymplectic diffeomorphisms acts transitivelly on $M$.
In fact, the local vector fields
$\displaystyle \left\{\derpar{}{x^i},\derpar{}{p_{i_1\ldots i_k}}\right\}$ are
locally Hamiltonian.
}
\end{remark}


\section{Invariance theorems}

(See \cite{EIMR-2012,Ib-2000} for more details).

As final remarks, in this Section we generalize some classical theorems 
of symplectic geometry in the field of multisymplectic manifolds.

The first one is a partial generalization of 
{\sl Lee Hwa Chung's Theorem} for symplectic manifolds,
which characterizes all the differential forms which are
invariant under infinitesimal symplectomorphisms \cite{Hw-47,LlR-88,GLR-84}:

\begin{theo}
\label{cgm}
Let $(M,\Omega )$ be a locally homogeneous multisymplectic manifold of degree $k$ and
$\alpha\in\df^p(M)$, with $p=k-1,k$, such that:
\begin{description}
\item{(i)}
The form $\alpha$ is invariant by the set of locally Hamiltonian
$(k-1)$-vector fields; that is,
$\Lie ({\bf X})\alpha =0$,  for every ${\bf X}\in\vf^{k-1}_{lh}(M)$.
\item{(ii)}
The form $\alpha$ is invariant by the set of locally Hamiltonian vector fields;
that is, 
$ \Lie (Z)\alpha =0$,  for every  $Z\in\vf_{lh}(M)$.
\end{description}
Therefore:
\ben
\item
If $p=k$ then $\alpha =c\, \Omega$, with $c\in\Real$.
\item
If $p=k-1$ then $\alpha =0$.
\een
\end{theo}
\noindent({\sl Outline of the proof\/} \cite{EIMR-2012}):
It is an adaptation of the proofs given in \cite{GLR-84,LlR-88}
for presymplectic and symplectic manifolds.
From the hypothesis of the Theorem and bearing in mind the properties
stated in Section \ref{sec3},  it can be proved that,
for every $X,Y\in\vf^{k-1}_{lh}(M)$, the following relation holds:
$\inn (X)\Omega\wedge\inn (Y)\alpha+\inn (Y)\Omega\wedge\inn (X)\alpha=0$;
and taking $X\in\ker^{k-1}\Omega$, from here you get to
$\inn (X)\alpha =0$.
Then it is proved that, if $p=k-1$ then $\alpha =0$;
but, if $p=k$, then there exists a unique $\alpha'\in\Cinfty(M)$ such that
$\inn (X)\alpha =\alpha'\inn (X)\Omega$,
for every $X\in\vf^{k-1}_{lh}(M)$.
Therefore, using some local properties of the locally Hamiltonian
$(k-1)$-multivector fields, it is concluded that $\alpha'$ is constant
and the final conclusion follows straightforwardly from the last results and Proposition \ref{propo3}.
\qed

The second one is a  generalization of some {\sl Theorems of
 Banyaga}
for symplectic and other orientable manifolds \cite{Ba86}:

\begin{theo}
Let $(M_i,\Omega_i)$, $i=1,2$, be local homogeneous
multisymplectic manifolds of degree $k$ and
$G(M_i,\Omega_i)$ their groups of multisymplectic automorphisms.
Let $\Phi\colon G(M_1,\Omega_1)\to G(M_2,\Omega_2)$
be a group isomorphism
(which is a homeomorphism when $G(M_i,\Omega_i)$ are endowed with the point-open topology).
 Then, there exists a diffeomorphism
$\varphi\colon M_1\to M_2$, such that :\begin{enumerate}
\item
$\Phi (\psi) = \varphi \circ \psi\circ\varphi^{-1}$,
 for every $\psi\in G(M_1,\Omega_1)$.
\item
The map $\varphi_*$
maps locally Hamiltonian vector fields of $(M_1,\Omega_1)$
into locally Hamiltonian vector fields of $(M_2,\Omega_2)$.  
\item
In addition, if $\varphi_*$ maps 
locally Hamiltonian multivector fields of $(M_1,\Omega_1)$
into locally Hamiltonian multivector fields of $(M_2,\Omega_2)$,
 then there is a constant $c$ such that
 $\varphi^*\Omega_2 = c\,\Omega_1$.
\end{enumerate}
\end{theo}
\noindent({\sl Outline of the proof\/} \cite{EIMR-2012}):
By Theorem \ref{theor1}, $G(M_i,\Omega_i)$ acts transitively on
$M_i$ and, by the main theorem in \cite{We54},
there exists a bijective map $\varphi\colon M_1\to M_2$ such that 
$\Phi (\psi) = \varphi \circ\psi\circ\varphi^{-1}$.  
Then it is proved that $\varphi$ is a homeomorphism and,
adapting the proof in \cite{Ba86} to our setting,
that it is also a smooth diffeomorphism.
Therefore, as a consequence of this proof, we conclude that
 $\varphi_*$ maps locally Hamiltonian vector fields
into locally Hamiltonian vector fields.
Finally, assuming the hypothesis of the third item,
using Theorem \ref{cgm} we have that
$\varphi^*\Omega_2 = c\, \Omega_1$.
\qed



\section{Conclusions and discussion}

Some of the main properties and characteristics of multisymplectic manifolds
have been reviewed in this disertation.
Although most of them are generalizations
of other well-known results for symplectic geometry,
in the multisymplectic case, they are more elaborated and richer than for symplectic manifolds, in general;
and it is for this reason that this is a topic of active research \cite{RW-2018}.

In particular, other interesting properties of multisymplectic manifolds
which have not been analyzed here are, for instance:
the {\sl graded Lie algebra structure} of the sets of {\sl Hamiltonian forms}
and {\sl Hamiltonian multivector fields} \cite{Ca96a,EIMR-2012},
{\sl polarized multisymplectic manifold} and its general structure theorem \cite{Ca96b},
as well as
other properties and relevance of $r$-coisotropic, $r$-isotropic
and, especially, of $r$-Lagrangian distributions and submanifols
\cite{Ca96b,LMS-2003},
and the characterizations of multisymplectic transformations \cite{EIMR-2012}.


\section*{Acknowledgments}

I acknowledge the financial support of the 
{\sl Ministerio de Ciencia e Innovaci\'on} (Spain), projects
MTM2014--54855--P and
MTM2015-69124--REDT,
and of {\sl Generalitat de Catalunya}, project 2017--SGR--932.


\end{document}